\documentclass[twocolumn,superscriptaddress,showpacs,preprintnumbers,amsmath,amssymb,prb]{revtex4-1}

\usepackage{verbatim}
\usepackage{epsfig}
\usepackage{graphicx}
\usepackage{color}
\usepackage[colorlinks,bookmarks=false,citecolor=blue,linkcolor=red,urlcolor=blue]{hyperref}
\usepackage{dcolumn}
\usepackage{bm}

\begin{document}

\title{Vortex Core Ordering in Abrikosov Lattices}

\author{Madhuparna Karmakar}
\email{madhuparna.k@gmail.com}
\affiliation{Department of Physics, Indian Institute of Technology Madras, Chennai 600036, India}
\author{R. Ganesh}
\email{ganesh@imsc.res.in}
\affiliation{The Institute of Mathematical Sciences, HBNI, C I T Campus, Chennai 600 113, India}

\begin{abstract}
  Vortex lattices, arising from repulsive inter-vortex interactions, are a canonical example of emergent phenomena. Recent studies on the cuprates have drawn attention to the appearance of competing correlations within vortex cores. This gives internal structure to each vortex and changes the nature of inter-vortex interactions, especially at short range. This can have significant consequences for vortex lattices wherein vortices are in close proximity to one another. We consider the attractive Hubbard model, a prototypical model for phase competition, with checkerboard charge density wave order appearing in vortex cores. We use next-nearest-neighbour hopping, $t'$, to tune the strength of competing order in the core. Using Bogoliubov deGennes mean field simulations, we study the variation in quasiparticle gap, superfluid stiffness and shear stiffness, as the core order is tuned. The presence of competing order allows for a new defect: domain walls that separate regions with different types of charge order overlying the vortex-laden superfluid background. We find the energy cost of such domain walls. Our results indicate a stable intermediate supersolid region with coexisting superfluid and charge orders. At one end, this phase is unstable to melting of the vortex lattice into a purely charge ordered state, possibly via an intervening liquid-like state. At the other end, charge order becomes incoherent due to domain formation while superconducting order persists.
\end{abstract}  
 
\pacs{}
\maketitle

\section{Introduction}
Vortices are topological defects in superconductors and superfluids, characterized by a winding of the complex order parameter. As their topological character endows them with stability, vortices can be thought of as `emergent' particles. This is best illustrated in type-II superconductors (and in superfluids with induced angular momentum) when vortices arrange themselves into regular patterns to form an `Abrikosov lattice'\cite{Abrikosov2004}. The formation, stability and response of these lattices is a long standing problem at the interface of soft and hard condensed matter physics\cite{Sonin1987,blatter_rmp}.

Recent experiments on cuprate superconductors have opened a new dimension to vortex studies. It has been demonstrated that vortex core regions can support competing phases, particularly charge order\cite{vortexcoreAFM,Lake2001,Chang2012,Wu2013,Machida2015,Yu2016,Ghiringhelli2012,Yoshizawa2013}. While superconducting pairing may overwhelm a competing order elsewhere, it cannot do so in the vortex core region where pairing itself must necessarily vanish. Several theoretical studies have examined these issues. In particular, many have taken a field theoretic approach using an enlarged local order parameter space subject to a constant amplitude constraint\cite{Meier2013,Wachtel2015,Morice2017}. 

In this article, we examine the effect of ordered cores on the macroscopic properties of the vortex lattice.
We may expect to see new physics here due to changes in the nature of inter-vortex interactions. The study of inter-vortex interactions in superconductors and superfluids has a long history\cite{Jacobs1979,Chaves2011}. In typical type-II superconductors, it is known that vortices repel one another, and thereby stabilize a lattice arrangement. This must be revisited when competing orders manifest within vortex cores, a situation that was first discussed within the hypothesized SO(5) theory for the cuprates\cite{Arovas1997}. This can induce an additional short-ranged attractive interaction between vortices\cite{Mackenzie2003}. 
In an earlier work, we demonstrated the occurrence of charge-density-wave-ordered cores in the attractive Hubbard model\cite{Karmakar2017}. This offers an ideal toy system that is amenable to microscopic studies. In this article, we build upon our earlier work: we tune the strength of core-ordering and study the effect on vortex lattice properties such as stability to shear. 
 
The occurrence of order in the cores also introduces new degrees of freedom in the description of vortex lattices. It can lead to independent ordering/disordering phenomena that are not tied to superconductivity. 
We demonstrate this in a situation where each vortex has $\mathbb{Z}_2$ character, due to two possible charge orderings. In a dense vortex lattice, this degree of freedom is ferromagnetically coupled between vortices, leading to an effective Ising ferromagnet defined on top of the vortex lattice. The co-occurrence of superconductivity and Ising ferromagnetism, both long-ranged, constitutes `supersolidity'.
New phenomena can emerge as a result, e.g., we find domain walls of the Ising field. Propagation of such domain walls destroys the competing $\mathbb{Z}_2$ order, while leaving superconductivity intact.

\section{Model and method}
The attractive Hubbard model is a prototypical model to describe phase competition and vortex core order. It describes fermions on a square lattice with on-site attraction,
\begin{eqnarray}
\nonumber H &=& \sum_{\langle ij \rangle,\sigma} \left[t_{ij} e^{i\theta_{ij}} c_{i,\sigma}^\dagger c_{j,\sigma} +h.c \right]-\mu \sum_{i,\sigma} \hat{n}_{i,\sigma} \\
&-& U\sum_i \hat{n}_{i,\uparrow} \hat{n}_{i,\downarrow}.
\label{eq.Hubbard}
\end{eqnarray}
Here, we take the hopping to be restricted to nearest neighbours with strength $t$ (set to unity) and next-nearest neighbours with strength $t'$. The complex phase of each hopping matrix element, $\theta_{ij}$, is set in accordance with the Peierl's substitution paradigm\cite{Karmakar2017}. We determine these phases such that we have a uniform magnetic flux that pierces the lattice, an approach suitable for strong type-II superconductivity. In this article, we confine our attention to $U=10$, the strong coupling case considered in Ref.~\onlinecite{Karmakar2017}.

At half-filling, with $t'=0$ and the magnetic field turned off, this model possesses an SO(3) symmetry with superconductivity (SC) and checkerboard charge-density-wave (CDW) orders becoming degenerate\cite{Yang1990,Burkov2008,Ganesh2009,Karmakar2017}. Next-nearest hopping, $t'$, serves as a tuning knob that lifts this degeneracy so that SC is energetically favourable over CDW order.
Upon introducing a magnetic field, we obtain vortices which manifest CDW correlations within their core regions\cite{Karmakar2017}.  

We study the properties of the Hamiltonian using a Bogoliubov deGennes (BdG) mean-field approach. Working on a $L\times L$ ($L=24$ below unless stated otherwise) square lattice with periodic boundary conditions, we define local SC and CDW order parameters at each site, $\Delta_i = U\langle c_{i,\downarrow} c_{i,\uparrow} \rangle$ and $\tilde{\phi}_i = \frac{U}{2} (-1)^{\mathbf{Q}\cdot \mathbf{r}_i} \left\{ \langle\hat{n}_{i,\uparrow} + \hat{n}_{i,\downarrow} \rangle-1\right\}$. We restrict all calculations here to half-filling by appropriately tuning the chemical potential, so that 
$ \left\{ \langle\hat{n}_{i,\uparrow} + \hat{n}_{i,\downarrow} \rangle-1\right\}$ represents local density deviation.
 The CDW order here is of the checkerboard type, characterized by density modulations at the wavevector $\mathbf{Q}=(\pi,\pi)$.
All order parameters are obtained self-consistently assuming zero temperature. The resolution in energy-per-site in our calculations is $\approx \frac{4t^2}{UL^2}$, as $\frac{4t^2}{U}$ is the relevant energy scale in a large-$U$ superexchange-like analysis. 

The CDW order in cores allows us to easily pin vortices using local delta function potentials. We take advantage of this below to calculate quantities of interest. We do this by generating suitably distorted mean-field solutions by using pinning potentials,
\begin{eqnarray}
H_{pinning} = \sum_i  W_i \{ \hat{n}_{i,\uparrow} + \hat{n}_{i,\downarrow}\}.
\label{eq.pinning}
\end{eqnarray}
The sites $i$ represent pinning locations. In our simulations on the attractive Hubbard model with $t'\lesssim0.5$, we find that the vortices are centred on sites of the square lattice, rather than on plaquettes. 
That is, each vortex has a minimum of the SC order parameter that occurs on a single site. 
As this vortex centre has a large density deviation due to CDW order, the vortex can be pinned by a local potential that couples to density. We achieve this by placing delta function potentials of strength $W_i$ at the required vortex positions. The sign of $W_i$, the potential, determines the nature of the CDW order, picking one of the two possible checkerboard orderings. 

We briefly summarize results from an earlier study\cite{Karmakar2017} here. For $t' \lesssim 0.5$, the Hubbard model of Eq.~\ref{eq.Hubbard} maps to an SO(3) Landau Ginzburg theory, with SC and CDW orders forming three components of a vector order parameter. The order parameters are coupled by a uniform length constraint whereby $\vert \Delta_i \vert^2 + \tilde{\phi}_i^2$ takes a fixed value at any position. In a superconducting vortex, SC order diminishes in the core while CDW order emerges to preserve the uniform length constraint. 
The resulting vortex profile has two length scales, albeit both related to one underlying length scale. Moving away from the core, the SC amplitude recovers to the asymptotic value over a distance $\xi_\Delta$ -- its profile is well approximated by a hyperbolic tangent curve. The CDW order attenuates over a longer length scale $\xi_\phi$ -- with a profile that is approximately a hyperbolic secant function. Both length scales ($\xi_\Delta$ and $\xi_\phi$) are inversely proportional to $t'$, the parameter which determines the cost of the competing CDW order. Generically, we have $\xi_\phi > \xi_\Delta$ so that CDW regions are larger than the SC vortex cores. To create vortices, we introduce an orbital magnetic field via the Peierl's substitution scheme in our BdG simulations. Due to periodic boundary conditions, a net magnetic flux can only be introduced in a quantized manner, with the value $\alpha h/e$, where $\alpha$ is an integer.  As each vortex carries a flux of $h/2e$, vortices are created in pairs. For a fixed $t'$ value, there is a critical magnetic flux strength beyond which CDW puddles around vortices overlap, giving rise to coherent long-ranged CDW order which co-exists with SC -- a `supersolid'.  

In  this article, we take a deeper look at the properties of the vortex lattice by keeping the magnetic flux constant at $\alpha=6$ (twelve vortices in the system) and tuning $t'$, the parameter which encodes asymmetry between CDW and SC orders. 
To keep the discussion simple, we restrict our attention to half-filling. We note that deviation from half-filling can serve as an independent tool to tune the competition between CDW and SC orders.
We examine four properties of the vortex lattice state: (a) quasiparticle gap, (b) superfluid stiffness, (c) shear stiffness, and (d) energy cost of a domain wall separating regions with differing CDW order. Our results shed light on the phases and phase transitions in the Hubbard model, and more generally, in systems with competing phases. 
 
\section{Inter-vortex interactions}

\begin{figure}
\includegraphics[width=3.2in]{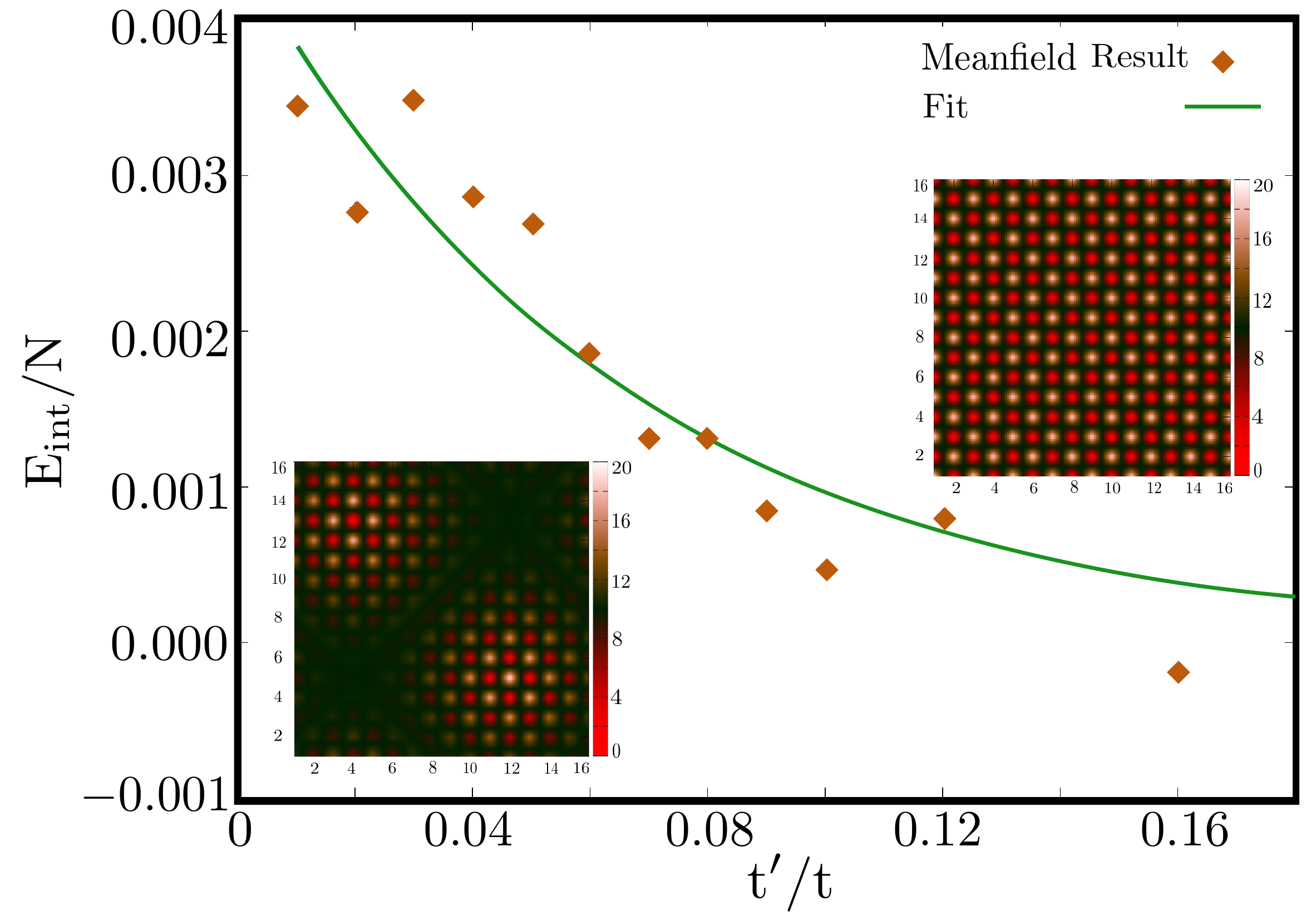}
\caption{Color online: Vortex-vortex interaction due to CDW core order. The Y axis shows the energy difference between two-vortex configurations with same (upper right inset) and opposite (lower left inset) CDW character. 
The insets show maps of self-consistently obtained density values, with CDW correlations visible.
The fit to an exponential function is shown. Energies are obtained from simulations on a 16$\times$16 lattice. The energy-per-site resolution is $\sim 0.001 (\approx 4t^2/16^2 U)$. 
}
\label{fig.Eint}
\end{figure}

In a type-II superconductor, vortices are known to repel one another. The occurrence of CDW order in cores is likely to modify the nature of the interaction. At high vortex densities, the vortices are located so close to one another that the CDW halo surrounding one vortex overlaps with that of each adjacent vortex. If the vortices possess the same sense of CDW ordering, this leads to coherent CDW formation in the region between the vortices. This can lead to an attractive interaction, as the energy gain from CDW formation would be higher if the vortices were closer. 
A quantitative demonstration of this effect has been given in the SO(5) model suggested for the cuprates\cite{Mackenzie2003}, albeit with the assumption of well-separated vortices. As the attractive Hubbard model of Eq.~\ref{eq.Hubbard} has been shown\cite{Karmakar2017,Karmakar2017b} to map to an SO(3) theory that is closely analogous to the SO(5) theory, this result is also applicable here. 
We hypothesize that coupling in the CDW sector leads to an additional interaction between two vortices, given by 
\begin{eqnarray}
\nonumber H_{v-v,CDW}(\mathbf{r}_i-\mathbf{r}_j) &\sim& -w_i w_j \exp\left\{ -\vert \mathbf{r}_i-\mathbf{r}_j \vert/\xi_\phi \right\} \\
&\equiv& -w_i w_j A\exp\left\{ -B\vert \mathbf{r}_i-\mathbf{r}_j \vert t' \right\}. \phantom{ab}
\label{eq.intform}
\end{eqnarray} 
Here, $w_i, w_j = \pm1$ are $\mathbb{Z}_2$ quantities that represent CDW character in the vortex cores. They indicate whether a given sublattice of the underlying square lattice has higher or lower density. If two vortices have CDW order of the same sense, they are subject to an attractive potential that decays over a length scale set by $\xi_\phi$, the size of the CDW cloud surrounding a single vortex. As $\xi_\phi$ scales as $1/t'$\cite{Karmakar2017}, we may have the above functional dependence on $t'$. 
The exponential form of the interaction may be rationalized as follows. The CDW order parameter profile in a vortex core is well approximated by the hyperbolic secant function. At long distances, it represents exponential decay. With two adjacent vortices, the overlap of their CDW profiles contributes to the interaction energy. As the CDW profiles decay exponentially, the CDW-contribution to the interaction will also take an exponential form.  

To verify the interaction form given in Eq.~\ref{eq.intform}, we obtain solutions with two vortices ($\alpha=1$). The vortices occur at a separation of $L/\sqrt{2}$, the maximum possible separation on an $L\times L$ lattice. We compare two solutions: (a) with the same CDW order at both vortices, and (b) with opposite CDW ordering (see insets to Fig.~\ref{fig.Eint}). To obtain these solutions, we use suitable pinning potentials as in Eq.~\ref{eq.pinning}. Energies of these solutions are calculated with respect to the Hamiltonian of Eq.~\ref{eq.Hubbard}, without pinning potentials. The difference, $E_{int} = E_{(b)}-E_{(a)}$, is a measure of the inter-vortex interaction in the CDW sector. Our results on a $16 \times 16$ lattice are shown in Fig.\ref{fig.Eint}. Although choosing $L=16$ leads to poorer energy resolution, it allows for a shorter separation between vortices, giving an interaction strength that can be ascertained within our resolution. We find a reasonable fit to an exponential in $t'$, consistent with Eq.~\ref{eq.intform}. 

Lattice arrangements of vortices, or even atoms, owe their stability to repulsive inter-particle interactions. Each particle sits at a minimum of the potential landscape created by repulsion from neighbouring particles. In vortices with competing order, as argued above, core-ordering leads to an attractive component in the inter-vortex interaction. This effectively weakens the repulsion between vortices and softens the potential minima that occur at lattice sites. Vortices can then become easily dislodged and mobile. This suggests a possible `quantum crystal' phase, analogous to liquid Helium\cite{Andreev1969}. A resulting melting of the vortex lattice will give rise to an incoherent superconductor, akin to a quantum spin liquid. At the mean field level, this can manifest as vanishing shear stiffness, as we discuss below.
 
\section{Competition between orders}
\begin{figure}
\includegraphics[width=3.2in]{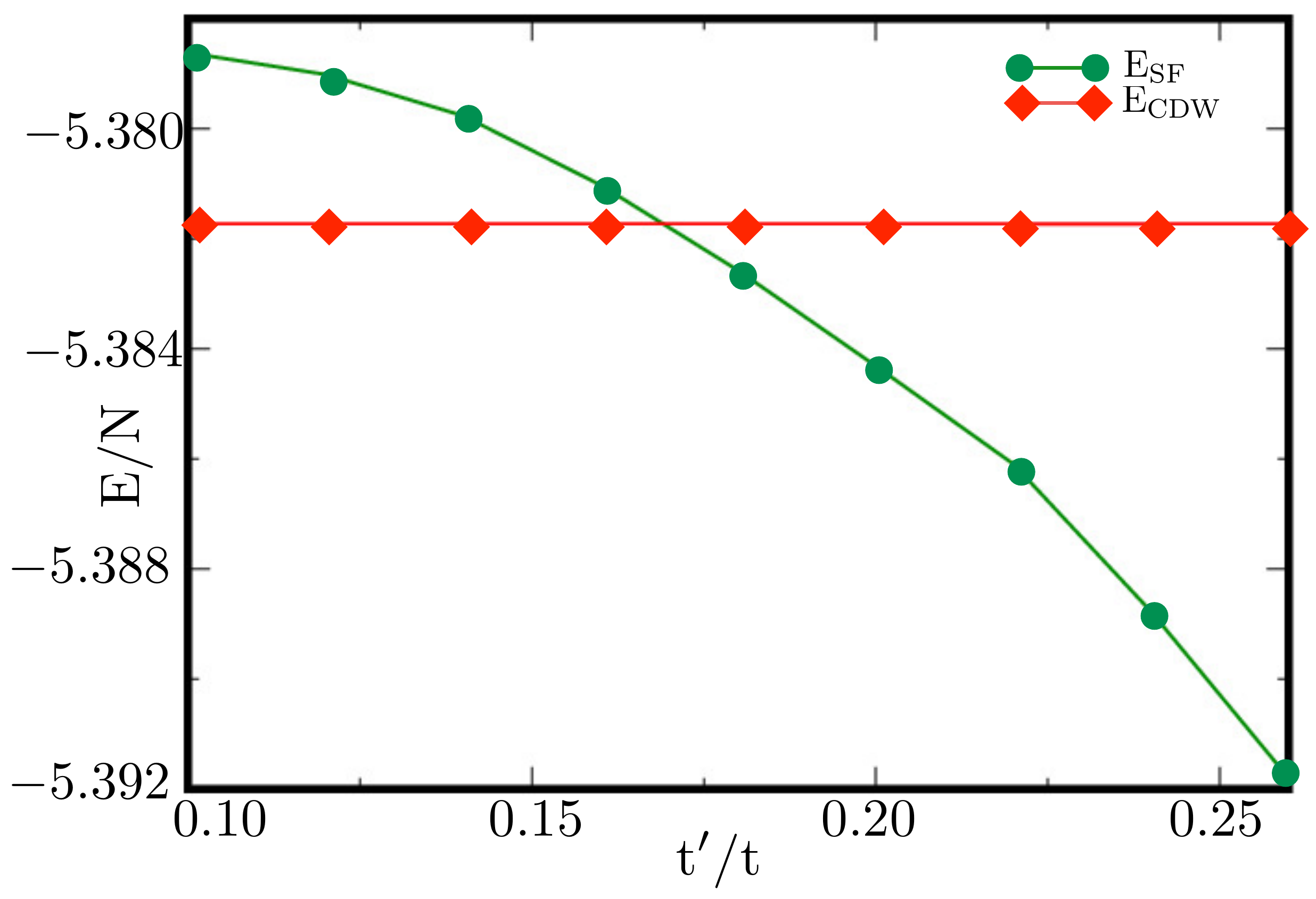}
\caption{Color online: Competing superconductivity and CDW orders. Energies of the uniform CDW state and the vortex-laden SC state are plotted vs. $t'$. The magnetic flux is kept fixed at $\alpha=6$, corresponding to twelve vortices piercing the $24\times 24$ lattice in the SC phase. With $L=24$, the energy-per-site resolution is $\sim5 \times 10^{-4} (\approx 4t^2/24^2 U)$. 
}
\label{fig.Ecomp}
\end{figure}

Competition between SC and CDW orders manifests as charge ordering within vortex cores. This is obtained from solutions of the BdG equations with SC order, but with local CDW correlations within vortex cores. However, the BdG equations also have other solutions, in particular, a uniform CDW solution with an insulating quasiparticle gap. Upon switching off the $t'$ hopping and the magnetic field, this CDW solution is degenerate with the uniform SC solution. The $t'$ hopping raises the energy of the CDW state. The applied field, however, does not change its energy as it is insulating. The SC state, in contrast, develops strong order parameter gradients in a field due to vortex formation. Fig.~\ref{fig.Ecomp} compares the energy of the uniform CDW state with that of the vortex-laden SC state as a function of $t'$, keeping the magnetic flux fixed at $\alpha=6$, corresponding to twelve vortices piercing the SC. 
At large $t'$, the SC state is energetically favourable even though it suffers from large order parameter gradients. At low $t'$ values, the CDW state wins. At $t'\sim0.17$, we see a crossing in the energies. Naively, this signifies a first order phase transition from SC to CDW orders. 
While we do not have a SC ground state for $t' \lesssim 0.17$, we do nevertheless find metastable SC solutions in this regime. 

\begin{figure}
\includegraphics[width=3.2in]{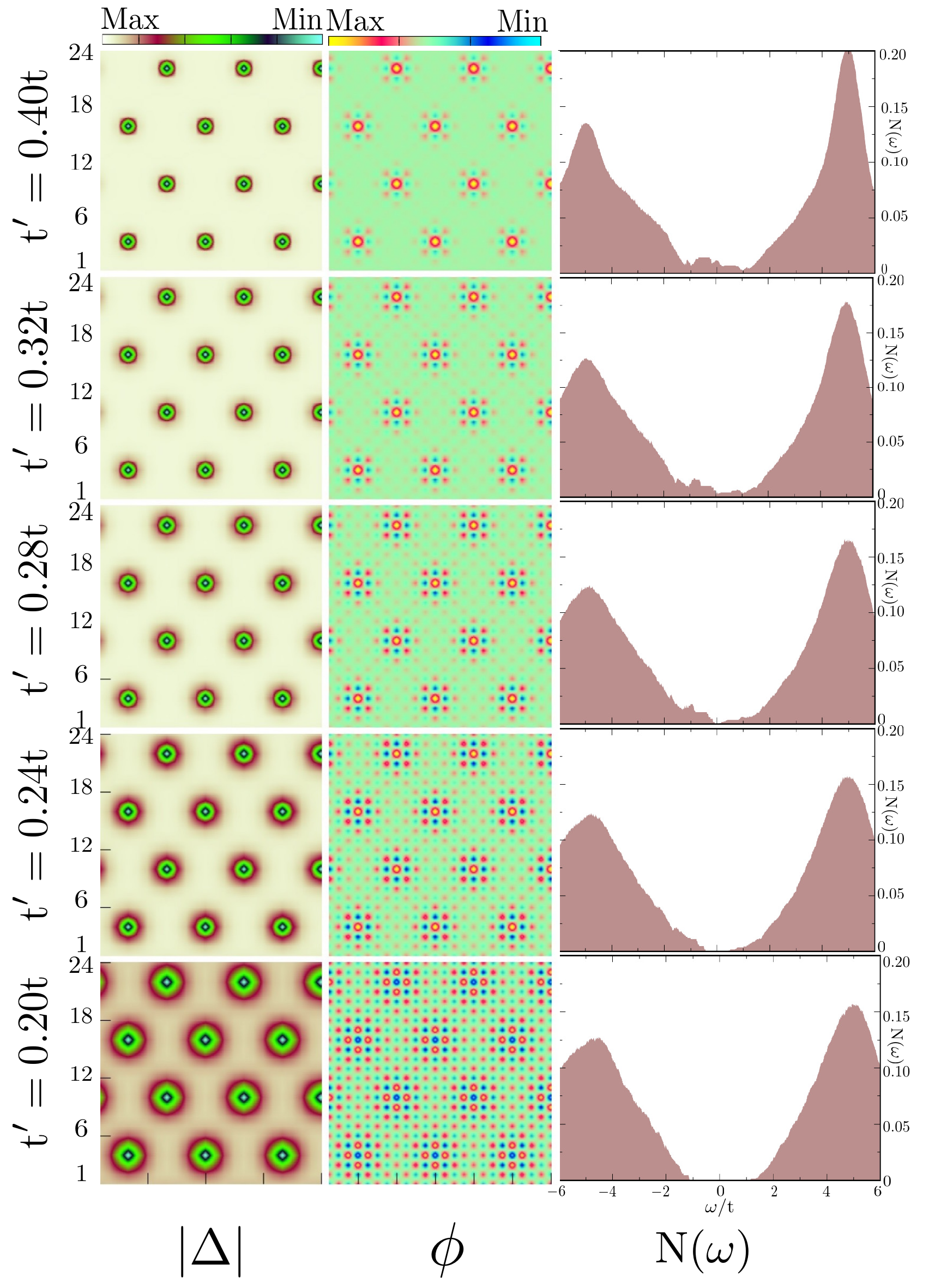}
\caption{Color online: Order parameters and DOS as for different $t'$ values, with magnetic flux fixed at $\alpha=6$. The left column shows the amplitude of the superconducting order parameter, with twelve vortices visible in each panel. The central column shows the self-consistently obtained density, with CDW order visible within vortex cores. The right column shows the quasiparticle DOS. There is  no quasiparticle gap for $t'\gtrsim 0.32$.}
\label{fig.maps}
\end{figure}

\section{Quasiparticle gap}
Both SC and CDW orders, if they occur in a uniform fashion, open gaps in the single particle spectrum. 
The SC state, in the presence of a magnetic field, necessarily develops strong gradients due to vortex-formation. As CDW order appears in vortex cores, this state also has gradients in the CDW order parameter. These gradients modify the quasiparticle gap. 

Fig.~\ref{fig.maps} shows the order parameters as a function of $t'$. It also shows the quasiparticle density of states (DOS). For small $t'$ values, we see a clear gap in spectrum. In this regime, CDW order is spread throughout the system. Indeed, the CDW order parameter does not go to zero at any point. This long ranged order succeeds in opening a gap. The gap decreases with increasing $t'$ and vanishes at $t' \gtrsim0.32$. At this point, CDW clouds around vortices no longer overlap. As a result, both CDW and SC orders have strong gradients, with neither order parameter remaining non-zero everywhere. This leads to states with energy close to zero, which turn out to be sharply localized at edge centres and plaquette centres of the vortex lattice -- saddle points of the order parameter profiles. 

The closing of the gap requires large fields and weak competing order (large $\alpha$ and large $t'$). In an earlier study restricted to $t'=0.2$, we always found a robust gap\cite{Karmakar2017}. The occurrence of zero energy states due to strong gradients is reminiscent of the effect of randomly placed impurities in the Hubbard model\cite{Karmakar2017b}. 

\section{Superfluid stiffness}
The stability of superconductivity is reflected in the energy cost of flow, i.e., an imposed phase winding. We estimate this quantity in the following way. We introduce an additional component of the vector potential, $\vec{A}_{winding} = 2\pi \hat{x}/L$. This does not lead to a flux perpendicular to the lattice plane. Rather, it induces an additional phase twist in the SC order parameter from one end of the lattice to another. Upon incorporating this additional vector potential, we find a `winding solution' to the BdG self-consistency equations. We use this solution in the original Hamitonian (without $\vec{A}_{winding}$) to find the energy of the flowing state, $E_{flow}$. The superfluid stiffness is obtained as $\rho_s = E_{flow}-E_0$, where $E_0$ is the energy of the non-flowing BdG solution (solution obtained without $\vec{A}_{winding}$).

\begin{figure}
\includegraphics[width=3.2in]{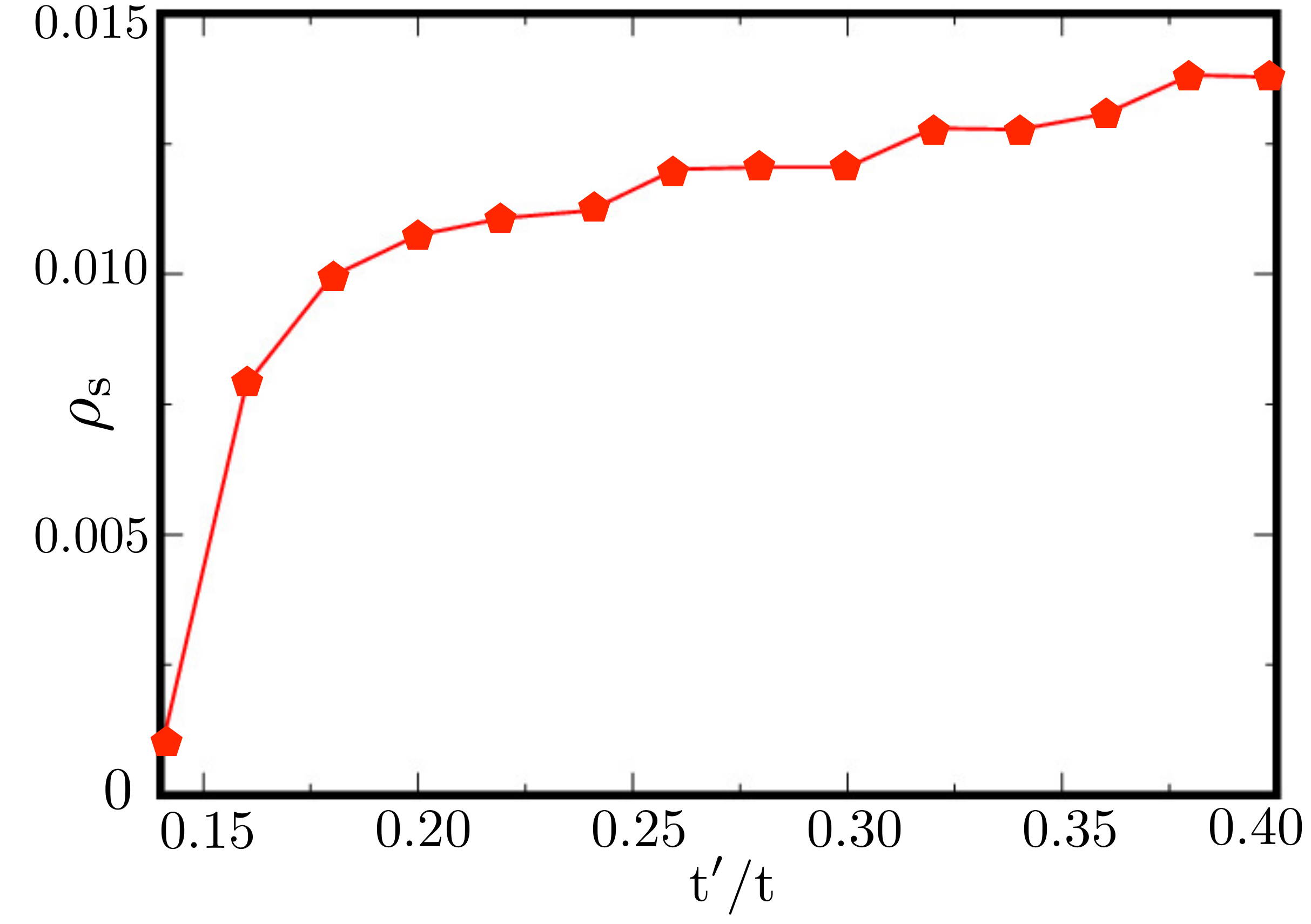}
\caption{Color online: Superfluid stiffness vs. $t'$ with magnetic flux fixed at $\alpha=6$.}
\label{fig.sfstiffness}
\end{figure}

Our results for superfluid stiffness are shown in Fig.~\ref{fig.sfstiffness}. We find that the superfluid stiffness increases with increasing $t'$. As $t'$ increases, CDW order is suppressed and SC order is strengthened, leading to increasing stiffness. The non-zero value of stiffness indicates robust SC correlations for any $t' \gtrsim 0.17$, the region where the SC state has lower energy than a uniform CDW solution (see Fig.~\ref{fig.Ecomp}).   
This result should be understood as being appropriate for weak disorder that pins vortices, but does not change the mean-field equations significantly. 
This is because a clean vortex system is expected to have zero stiffness, due to motion of vortices along an applied field. Such dynamical effects lie beyond our mean-field approach. 
Even if disorder were so weak as to only pin a single vortex, the vortex lattice provides additional rigidity as all other vortices lock into a lattice arrangement. 

\section{Shear stiffness}

\begin{figure*}
\includegraphics[width=6.6in]{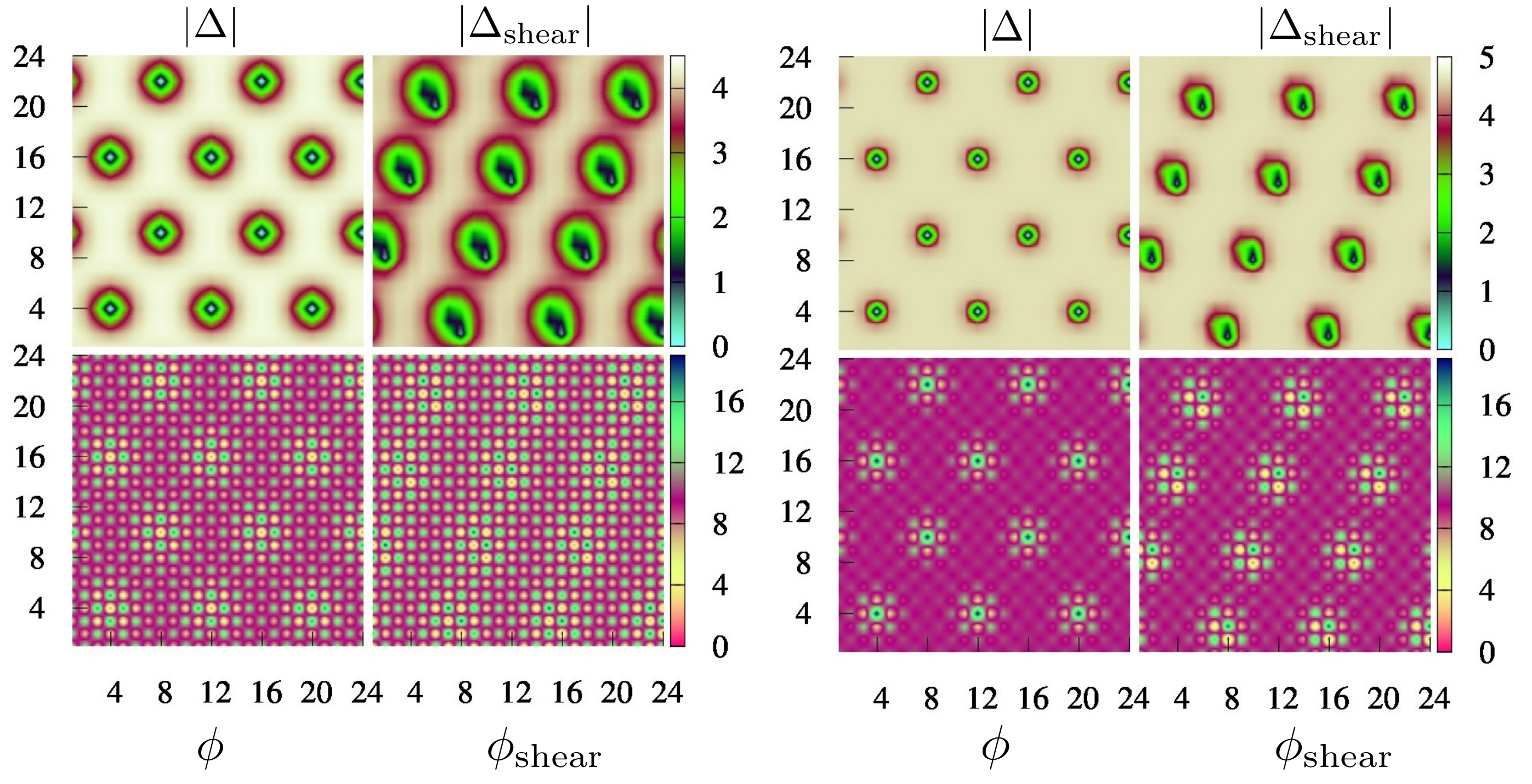}
\caption{Color online: Sheared vortex lattice for $t'=0.2$ (left) and $t'=0.34$ (right). The upper panels show the amplitude of the SC order parameter with and without shear. The lower panels show the self-consistent density, $\phi$, with CDW correlations clearly visible in vortex cores. }
\label{fig.sheared}
\end{figure*}

The natural way to quantify stability of a vortex lattice is via stiffness to shear. This is the energy cost incurred when the lattice parameters undergo an infinitesimal change.   
At a magnetic flux corresponding to twelve vortices on a $24\times 24$ lattice, we find a near-regular triangular arrangement of vortices for any $t'$ as shown in Fig.~\ref{fig.maps}. In order to shear the vortex lattice, we use pinning potentials to localize vortices on an oblique lattice, as shown in Fig.~\ref{fig.sheared}. While this is not an infinitesimal distortion of the vortex lattice, it is the minimal distortion that is commensurate with our finite size system. Shear appears to frustrate SC correlations, leading to weaker SC and stronger CDW order parameters. 

\begin{figure}
\includegraphics[width=3.2in]{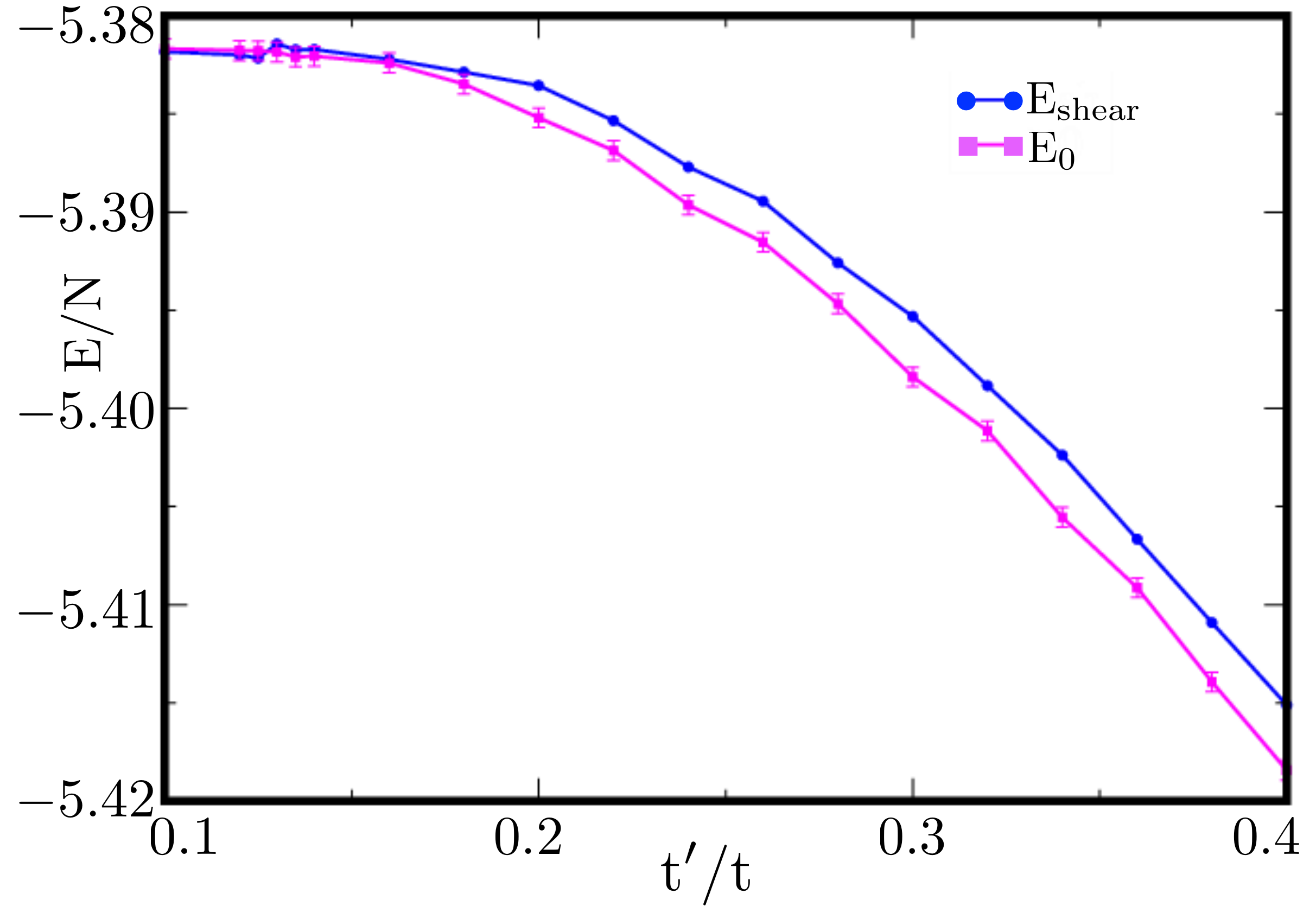}
\caption{Color online: Energy of sheared configuration vs. that of lowest energy solution. The error bars in $E_0$ represent our energy resolution on a $24\times 24$ lattice. Within this resolution, energy cost of shear vanishes at $t'\sim$ 0.18.}
\label{fig.Eshear}
\end{figure}

To find the energy cost of shear, we first obtain self-consistent SC and CDW order parameters with pinning potentials placed at sheared lattice positions. We take this solution and calculate its energy in the original Hamiltonian of Eq.~\ref{eq.Hubbard} (without pinning potentials).
Our results are shown in Fig.~\ref{fig.Eshear}. We find that shear stiffness is non-zero for $t' \gtrsim 0.18$. This indicates robustness of the vortex lattice at large $t'$, when CDW order is weak and the inter-vortex interaction is strongly repulsive. 
The critical $t'$ where shear stiffness vanishes lies very close to $t'\sim0.17$, where the energies of vortex-laden SC and uniform CDW states cross, as shown in Fig.~\ref{fig.Ecomp}. As we decrease $t'$, if shear stiffness indeed vanishes first, this signals a `quantum melting' transition into a spin-liquid-like state wherein vortices move about freely. This can be understood as a softening of the lattice potential due to inter-vortex attraction, which, in turn, arises from core-ordering. 
SC coherence will then be destroyed by mobile vortices. However, CDW order may still survive as we already have coherent long-ranged CDW order at this $t'$. This suggests the possibility of an exotic SC-CDW transition, with a possible intervening liquid-like state. 

\section{Domain wall energy}

\begin{figure*}
\includegraphics[width=6.6in]{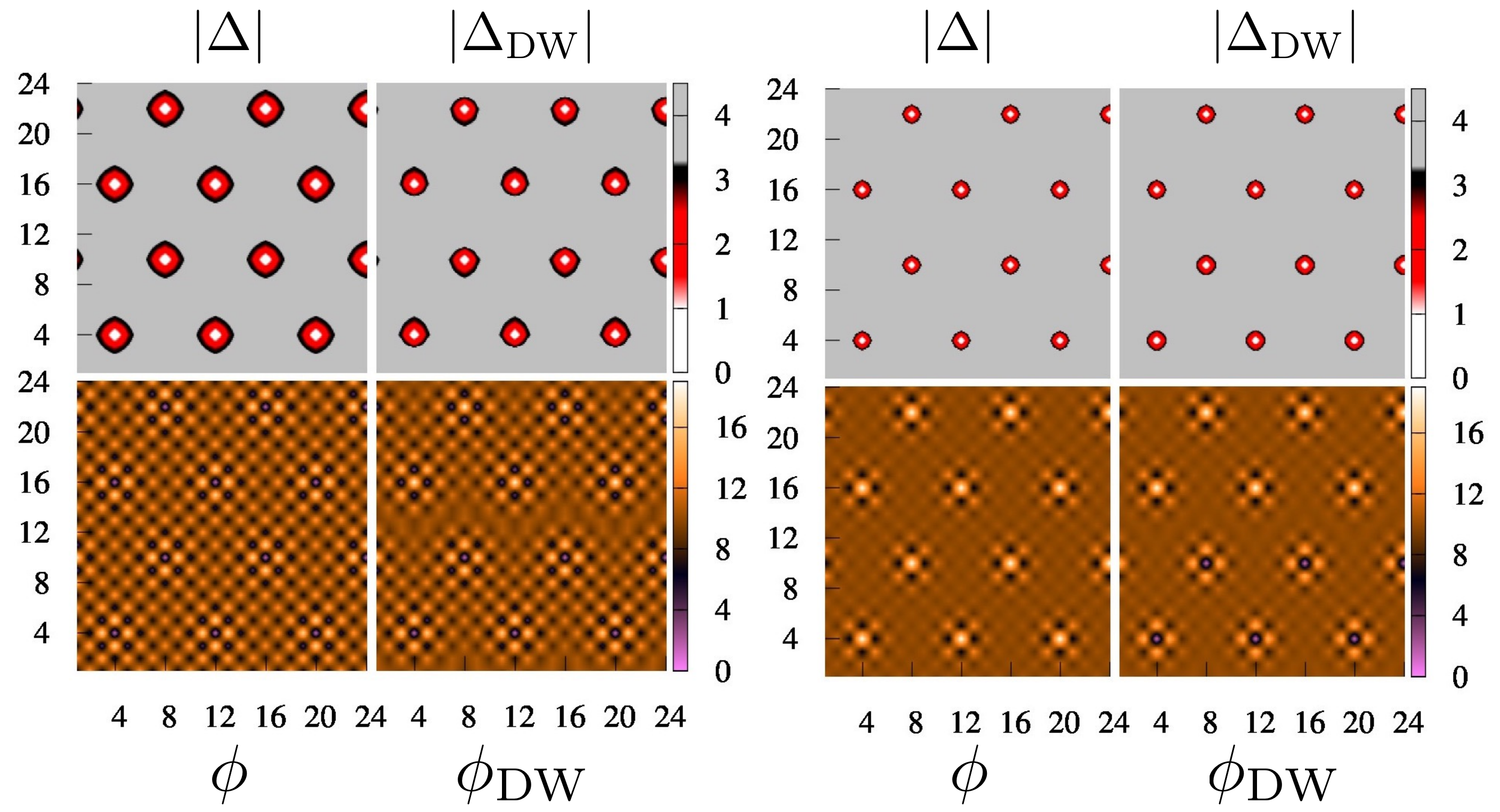}
\caption{Color online: SC amplitude and density with and without a domain wall for $t'=0.2$ (left) and $t'=0.34$ (right). The domain wall solutions have opposite CDW character in the lower and upper halves of the lattice. The vortex core positions have higher (lower) density in the upper (lower) half.
}
\label{fig.dw}
\end{figure*}
\begin{figure}
\includegraphics[width=3.2in]{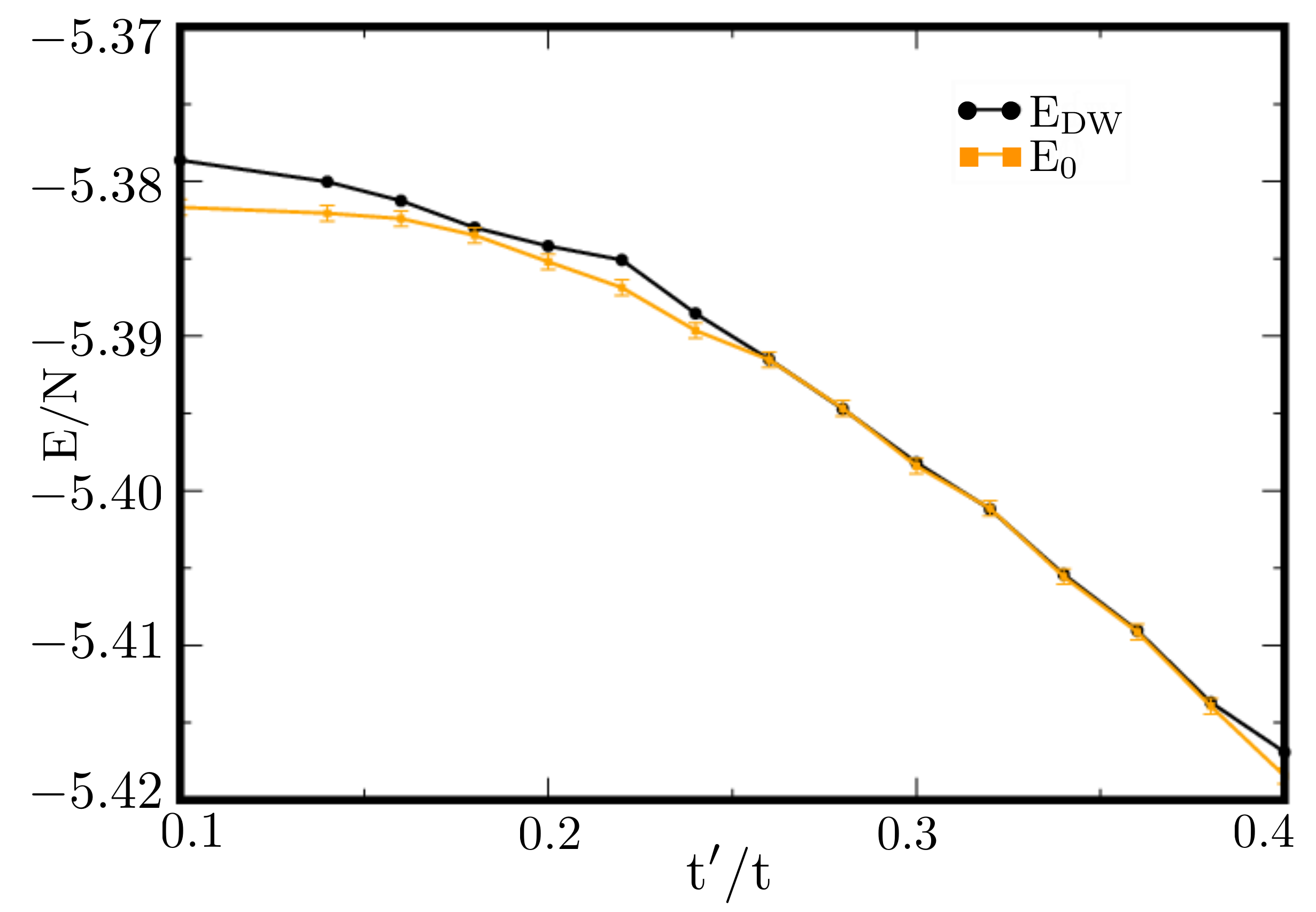}
\caption{Color online: Energy cost of a domain wall in CDW order vs. $t'$ with magnetic flux fixed at $\alpha=6$. The error bars in $E_0$ represent our energy resolution on a $24\times 24$ lattice. 
}
\label{fig.dwcost}
\end{figure}
The presence of competing order can lead to qualitatively new phenomena. The CDW order that develops on top of the SC background can independently develop defects, or even undergo a disordering transition. The CDW is described by an emergent Ising ferromagnet with moments at each vortex site. Each moment takes values $w_i = \pm 1$, corresponding to the two possible checkerboard orderings. For small $t'$, the CDW puddles around neighbouring vortices overlap strongly, leading to a ferromagnetic coupling between $w_i$'s. That is, it is energetically favourable to have similar CDW correlations at neighbouring vortex sites. As $t'$ increases, the overlap decreases, as does the Ising coupling. This suggests a $t'$-tuned Ising-like ferromagnet-paramagnet transition. 
As is well known in the two-dimensional Ising model, the paramagnetic transition is driven by the proliferation of domain walls. A reasonable estimate for the critical $t'$ value is that at which the mean-field energy cost of a domain wall vanishes.  

To obtain a domain wall solution, we use pinning potentials, as in Eq.~\ref{eq.pinning}, at the locations of vortices obtained from the lowest energy solution. We divide the lattice into two contiguous halves. We use a positive pinning potential for one half, and a negative potential for another. This leads to a vortex lattice as shown in Fig.~\ref{fig.dw}, with a domain wall in the CDW order. To find the energy cost of a domain wall, we take the converged solution with the pinning potential and evaluate its energy using the Hamiltonian of Eq.~\ref{eq.Hubbard} without any potentials. The difference of this energy from that of the usual (unpinned) mean-field solution gives the energy cost of a domain wall. 

The resulting energies are shown in Fig.~\ref{fig.dwcost}. We find that the domain wall cost vanishes for $t' \gtrsim 0.26$. This is the threshold at which CDW regions around vortices no longer overlap. This is consistent with the Ising coupling between neighbouring vortices going to zero. As a result, a domain wall does not cost any energy. We find a non-zero domain wall cost (within our resolution) for $0.18 \lesssim t' \lesssim 0.26$. In this intermediate region, we have long ranged CDW order due to strong overlap between neighbouring vortex cores. This corresponds to Ising ferromagnetic order with a robust magnetic moment. Surprisingly, at $t'\sim 0.18$, the domain wall cost once again vanishes. This roughly coincides with the point at which the shear stiffness vanishes. This suggests that the melting of the vortex lattice may also destroy CDW ordering. For $t'\lesssim 0.16$, the domain wall cost is once again non-zero. However, in this region, the vortex lattice is a metastable state as the uniform CDW state has lower energy as shown in Fig.~\ref{fig.Ecomp}.

\section{Discussion}
We have presented a mean-field study of the attractive Hubbard model in the presence of an orbital field. This model allows for CDW order to appear within the cores of superconducting vortices. We demonstrate that CDW order in the core induces an attractive component in the inter-vortex interaction. In a dense vortex lattice, the core regions overlap leading to percolation of CDW order throughout the system. This leads to `supersolidity', i.e., coexistence of density order and superconductivity as was first discussed in the context of Helium\cite{Boninsegni2012}. We study the stability of supersolid order, and that of the vortex lattice itself, as the strength of competing order is tuned via $t'$, the next-nearest neighbour hopping strength.

\begin{figure}
\includegraphics[width=3.2in]{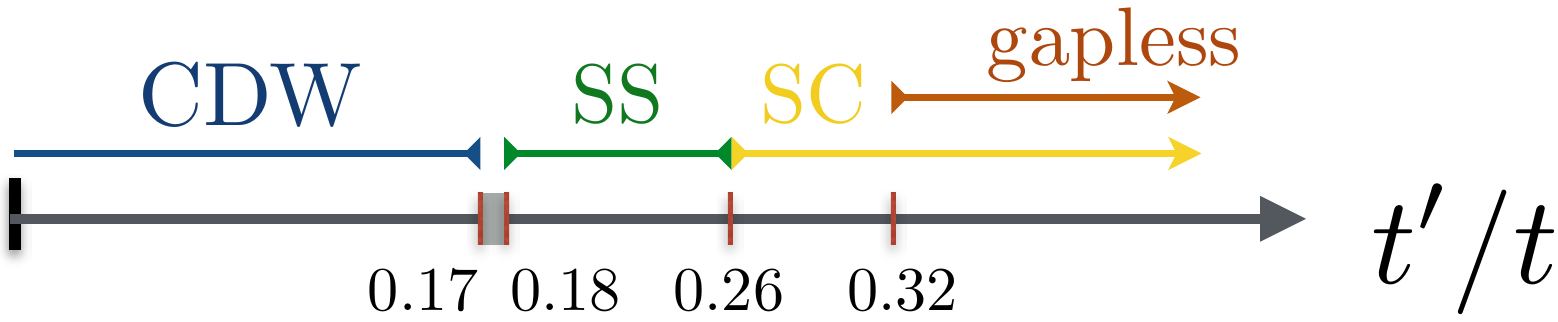}
\caption{Color online: Phase diagram as a function of $t'$, with magnetic flux fixed at $\alpha=6$ and interaction strength fixed at $U=10 t$. `SS' represents a supersolid state with coexisting long ranged superconductivity and CDW orders. `SC' represents a state with genuine superconductivity, but with no long-range CDW order. The small intervening region between CDW and SS regions may host a spin-liquid-like state resulting from the melting of the vortex lattice.
}
\label{fig.pd}
\end{figure}

Our results suggest a phase diagram as shown in Fig.~\ref{fig.pd}. 
Below $t' \sim 0.17$, we find a uniform CDW ground state with no SC character. For $0.18 \lesssim t' \lesssim 0.26$, we find a supersolid vortex lattice phase with non-zero values of quasiparticle gap, shear stiffness, superfluid stiffness and domain wall energy cost. 
In a narrow window at the CDW boundary, $0.17 \lesssim t' \lesssim0.18$, the shear stiffness and the domain wall cost vanish. This suggests melting of the vortex lattice due to quantum fluctuations, leading to a highly interesting spin-liquid-like state with incoherent SC and CDW orders.
At $t'\sim 0.26$, the domain walls soften while superfluid and shear stiffnesses remain non-zero. This suggests a quantum phase transition that is Ising-like: CDW order is lost while robust SC correlations persist. For $t' \gtrsim 0.32$, the superconductor becomes gapless as large gradients in the SC order parameter lead to localized states with energy close to zero. In Ref.~\onlinecite{Karmakar2017}, we had suggested a phase diagram for the attractive Hubbard model as $t'$ and the magnetic flux are varied. This work sharpens this phase diagram, providing clearer definitions of the phases and shedding light on the nature of the phase transitions.

We have used large scale real-space Bogoliubov deGennes mean field simulations, a method that has been widely used to study competing phases
\cite{takigawa2003,machida2005,balatsky2006,gong2004,machida2003,ting2009,machida2004,ghoshal2005,zhou2007,ting2003,hedegard2006}. 
This is the best available microscopic method to study vortex lattices, as the orbital field introduces a sign problem in quantum Monte Carlo simulations. While this approach is constrained by finite sizes, our work nevertheless highlights interesting features in the physics of supersolid vortex lattices. In particular, it suggests two interesting ordering phenomena: (a) melting of CDW order due to domain wall proliferation, while leaving SC correlations intact, and (b) melting of the vortex lattice itself for weak $t'$ values.  

Our work suggests new directions for experiments on cuprates. Recent experiments have uncovered charge order, showing that it is strengthened by the application of a magnetic field\cite{Wu2013,LeBoeuf2013}. Unlike the Hubbard model studied here, the charge order in the cuprates is incommensurate and not well understood. We have used $t'$ as a tuning parameter to tune the competition of SC with CDW order. In the cuprates, this role can be played by doping. Maximal charge ordering seems to occur at 1/8 doping. Moving away from this value, we may be able to tune the strength of vortex core order\cite{Grissonnanche2014,daSilvaNeto2016}. For large deviations in doping, charge ordering may disappear due to proliferation of defects -- analogous to the domain walls that we have discussed here. Such defects may always be present in the cuprates due to intrinsic disorder (a possible explanation for why coherent charge ordering has not been observed so far \cite{Achkar2014}). Nevertheless, it may be possible to see the melting of charge order by defect proliferation. 

Our arguments about inter-vortex interactions and melting may also have broader relevance. For example, multiband superconductors such as MgB$_2$ show complex inter-vortex interactions\cite{Lin2011}. This has given rise to a new classification of type-1.5 superconductors. These show interesting clustering phenomena that arise from softening of the vortex potential landscape\cite{Babaev2017}. Vortex interactions and ordering are also of great interest in experiments with ultracold atomic gases with synthetic vector potentials\cite{Dalibard2011,Kuno2015}. Our study of superconductivity-CDW competition in vortex lattices bears similarities to superfluid-Mott insulator competition in lattice bosonic systems\cite{Duric2010,Natu2016,Kuno2017}.  

\appendix

\subsection{Appendix: Bogoliubov-de-Gennes formalism}

We study the Hubbard model on a square lattice with periodic boundary conditions. Our Hamiltonian is given by 
\begin{eqnarray}
\nonumber H &=& \sum_{\langle ij \rangle,\sigma} \left[t_{ij} e^{i\theta_{ij}} c_{i,\sigma}^\dagger c_{j,\sigma} +h.c \right]-\mu \sum_{i,\sigma} \hat{n}_{i,\sigma} \\
&-& U\sum_i \hat{n}_{i,\uparrow} \hat{n}_{i,\downarrow} +  \sum_i  W_i \{ \hat{n}_{i,\uparrow} + \hat{n}_{i,\downarrow}\}.
\label{eq.Hubbardpin}
\end{eqnarray}
Here, $t$ (set to unity) and $t'$ represent the hopping strengths on nearest and next-nearest neighbour bonds. The hoppings are assumed to take complex phases $\theta$ (for nearest neighbours) and $\chi$ (for next-nearest neighbours) in accordance with the Peierls substitution paradigm. These phases ensure that a uniform magnetic field pierces the lattice. The lattice with periodic boundary conditions forms a torus; we fix the magnetic field to be that corresponding to six magnetic monopoles that are confined within the torus.
We include pinning potentials $W_i$ as described in the text, e.g., to study shear stiffness, we include pinning potentials at the sites of a sheared vortex lattice. Typically, we find that a pinning potential of strength $\vert W_i \vert \sim 1.0$ suffices to pin vortices to the required positions.

We decompose the on-site interaction term in pairing and density channels via a mean field 
decomposition. The complex superconducting order parameter is defined as, 
$\Delta_{i} = U\langle c_{i\downarrow}c_{i\uparrow}\rangle$, while the charge order parameter is defined as, 
$\phi_{i} = \frac{U}{2}(n_{i\uparrow}+n_{i\downarrow}) = \frac{U}{2}(\langle c_{i\uparrow}^{\dagger}c_{i\uparrow}\rangle + 
\langle c_{i\downarrow}^{\dagger}c_{i\downarrow}\rangle)$. The resulting effective Hamiltonian is given by
\begin{eqnarray}
\nonumber H_{MFT} & = & -t\sum_{\langle ij \rangle, \sigma}e^{i\theta_{ij}}c_{i\sigma}^{\dagger}c_{j\sigma}
 -t'\sum_{\langle\langle ij \rangle\rangle, \sigma}e^{i\chi_{ij}}c_{i\sigma}^{\dagger}c_{j\sigma}
 + h. c \\
  &+&\sum_{i,\sigma} \{W_i- \mu - \phi_i \} c_{i\sigma}^{\dagger} c_{i\sigma} \nonumber 
 - \sum_{i}(\Delta_{i}c_{i\uparrow}^{\dagger}c_{i\downarrow}^{\dagger} + \Delta_{i}^{*}c_{i\uparrow}c_{i\downarrow}) \\
 &+& \sum_{i} \{\vert \Delta_{i}\vert^2 + \phi_i^2\}/{\vert U \vert}.
\end{eqnarray}
To diagonalize this Hamiltonian, we use a Bogoliubov-Valatin transformation 
given by $c_{i\sigma}  =  \sum_{m}(u_{m i \sigma}\gamma_{m \sigma} - 
s_{\sigma}v_{m i \sigma}^{*}\gamma_{m, -\sigma}^{\dagger})$. Here, $\gamma_{m \sigma}^{\dagger}$
($\gamma_{m\sigma}$) creates (annihilates) a quasiparticle with spin $\sigma$ and 
energy $\epsilon_{m}^{\sigma}$. The quasiparticle wavefunctions are encoded by $u_{m i \sigma}$ and $v_{m i \sigma}$.
We have introduced a spin index $s_{\uparrow} = 1$ and $s_{\downarrow} = -1$.
The resulting gap and number equations are  
\begin{eqnarray}
 \Delta_{i} & = & U\sum_{m}\left\{v_{m i \downarrow}^{*}u_{m i \uparrow}f(\epsilon_{m\uparrow})
 + u_{m i \downarrow}^{*}v_{m i \uparrow}f(\epsilon_{m \downarrow})\right\}, \nonumber \\
 n_{i\uparrow} & = & \sum_{m}\left\{\vert u_{m i \uparrow}\vert^{2}f(\epsilon_{m \uparrow}) 
 + \vert v_{m i \uparrow}\vert^{2}f(\epsilon_{m\downarrow})\right\}, \nonumber  \\ 
 n_{i \downarrow} & = & \sum_{m}\left\{\vert u_{m i \downarrow}\vert^{2}(1 - f(\epsilon_{m\uparrow}))
 + \vert v_{m i \downarrow}\vert^{2}(1 - f(\epsilon_{m \downarrow}))\right\},\nonumber\\
\end{eqnarray}
where $f(\epsilon_m) = 1(0)$ if $\epsilon_{m}<0(>0)$ is the Fermi function at zero temperature.  
Starting from initial guesses, we iterate these equations to obtain self-consistent values of $\Delta_i$ and $\phi_i$. Using the number equations, we fix the value of the chemical potential to restrict the density to half-filling.
The mean-field energies that we obtain have a finite resolution due to system size effects. The resolution can be determined as the standard deviation in the energy upon varying the initial guess values. For the regime studied here, we find that the resolution in energy-per-site is approximately $4t^2/UL^2$.

\bibliographystyle{apsrev4-1} 
\bibliography{HubbardSO3,citations}
\end{document}